\documentclass[twocolumn,showpacs,preprintnumbers,floatfix,pre]{revtex4}

\usepackage{amsmath,amssymb}
\usepackage{graphicx}

\newcommand{\lm}{$\langle\lambda\rangle$ }
\newcommand{\am}{$\langle A\rangle$ }
\newcommand{\az}{$\varepsilon$ }

\begin{document}


\title{Granular spirals on erodible sand bed submitted to a circular fluid motion}

\author{H. Caps}
\author{N. Vandewalle}
\affiliation{GRASP, Institut de Physique B5, Universit\'e de Li\`ege, 
\\ B-4000 Li\`ege, Belgium.}
\begin{abstract}
An experimental study of a granular surface submitted to a circular fluid motion is presented. The appearance of an instability along the sand-water interface is observed beyond a critical radius $r_c$. This creates ripples with a spiral shape on the granular surface. A phase diagram of such patterns is constructed and discussed as a function of the rotation speed $\omega$ of the flow and as a function of the height of water $h$ above the surface. The study of $r_c$ as a function of $h$, $\omega$ and $r$ parameters is reported. Thereafter, $r_c$ is shown to depend on the rotation speed according to a power law. The ripple wavelength is found to decrease when the rotation speed increases and is proportional to the radial distance $r$. The azimuthal angle \az of the spiral arms is studied. It is found that \az scales with $h\omega r$. This lead to the conclusion that \az depends on the fluid momentum. Comparison with experiments performed with fluids allows us to state that the spiral patterns are not the signature of an instability of the boundary layer.
\end{abstract}


\maketitle

\section{Introduction}

Many experiments have been dedicated to the study of fluid flows between rotating stainless-steel disks \cite{jarre,hoffmann,gauthier,faller}. Three experimental configurations can be found in the literature: two contra-rotative disks; the upper disk rotating while the other is at rest; and the lower disk rotating, the other being at rest. For a given Reynolds number, i.e. at a given distance from the center of the plate \cite{schlichting}, a transition in the boundary layer occurs. The encountered instabilities are of K\`arm\`an, Ekman or B\"odewadt types and may lead to one of the following patterns: (i) concentric circles at a low rotation speed (ii) spirals at a medium rotation speed and (iii) a disordered state at large rotation speeds \cite{gauthier}. 

On the other hand, one can note the growing interest of the statistical physics community to the granular state. Particularly, the ripple formation on sand bed eroded by a fluid such as air or water has become a largely studied phenomenon \cite{frette,stegner,makse,thomas,farzam}. Despite the familiar aspect of the ripples, the physical mechanisms involved in their formation are related to complex phenomena of granular transport such as avalanches, saltation, reptation and suspension \cite{bagnold1}. 

In the present paper, both subjects will be combined in a single experiment. We will study the stability of a granular surface submitted to a circular fluid flow. In the next section, the experimental setup will be described. In Section III, we will present the results, and discuss them in Section IV. Finally, a summary of our findings is given in Section V.

\section{Experimental setup}

\begin{figure}[htb]
\begin{center}
\includegraphics[width=5cm]{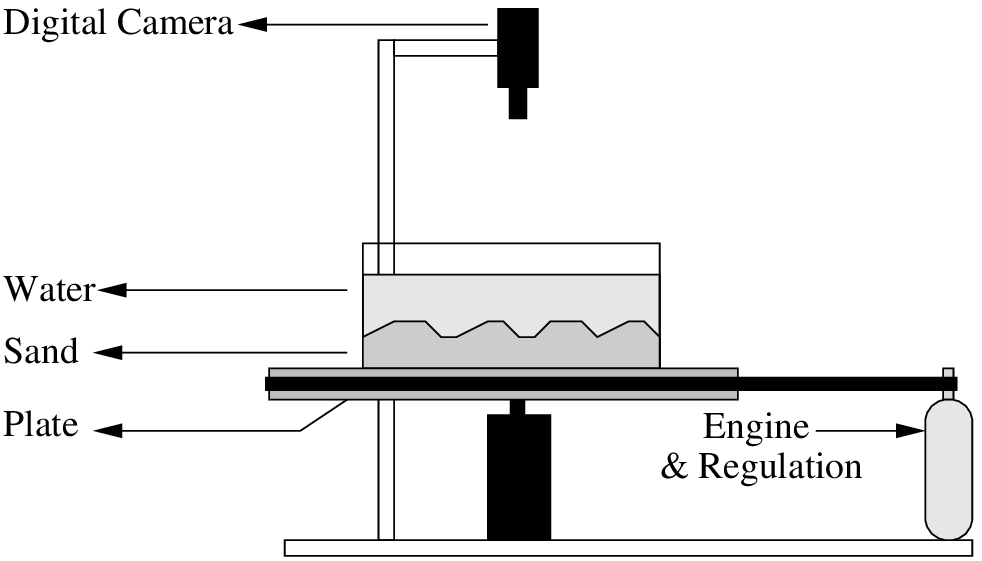}\\\vskip 0.2cm 
\includegraphics[width=8.5cm, height=0.7cm]{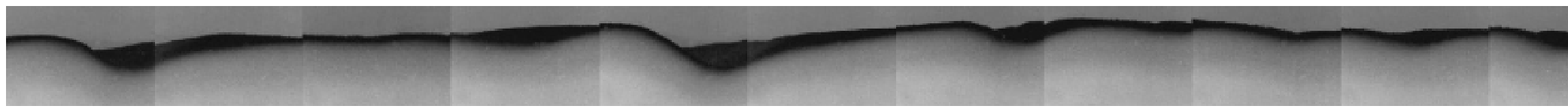}
\caption{(Top) Experimental setup for the ripple formation and 
analysis. (Bottom) Profile of the landscape along the circumference of 
the container after the formation of the ripples. Notice that the 
image has been extended vertically for clarity.}\label{manip}
\end{center}
\end{figure}

Our experimental setup is illustrated in Fig.\ref{manip} and consists of a horizontal circular plate connected by a belt to an engine. The rotation speed can be adjusted from 8 rpm up to 100 rpm. A cylindrical container (radius $R=11$ cm) filled with water and sand (mean grain diameter $d=282.5\ \mu$m) is placed in the center of the plate and put into rotation. When the rotation is brutally stopped, the sand bed remains fixed while the water continues its inertial circular motion. The shear stress applied to the sand by the water flow initiates various grain motions: saltation, reptation and  suspension. After a short time (typically 2s), current ripples are formed [see Fig.\ref{manip} bottom].  

A CCD camera is placed on the top of the plate and records an image of the sand bed. In order to avoid shadow effects, the container circumference is homogeneously illuminated. An annular luminescent tube with a radius of 12 cm is placed horizontally around the container, at the level of the sand-water interface. We assume that the height of sand is given by the intensity of the pixels on the grayscale images, due to this side lightning. Within this assumption, the landscape is given with an accuracy of $0.7\,d\approx200\ \mu m$.

The parameters of our experiment are: (i) the height of water $h$ over the sand bed before rotation, and (ii) the rotation speed $\omega$ of the container. Moreover, one should note that the Reynolds number varies on the granular surface as a function of the distance $0<r<R$ from the center of the container \cite{schlichting}. Therefore, the Reynolds number $R_e$ we will use can be written as
\begin{equation}\label{reynolds}
R_e=\sqrt{r^2\frac{\omega}{\nu}},
\end{equation}
where $\nu$ is the kinematic viscosity of the water, i.e. 
$\nu=10^{-6}$ m${}^2$ s${}^{-1}$. 

\section{Results}

\begin{figure}[htb]
\begin{center}
\includegraphics[width=8cm, 
height=1.5cm]{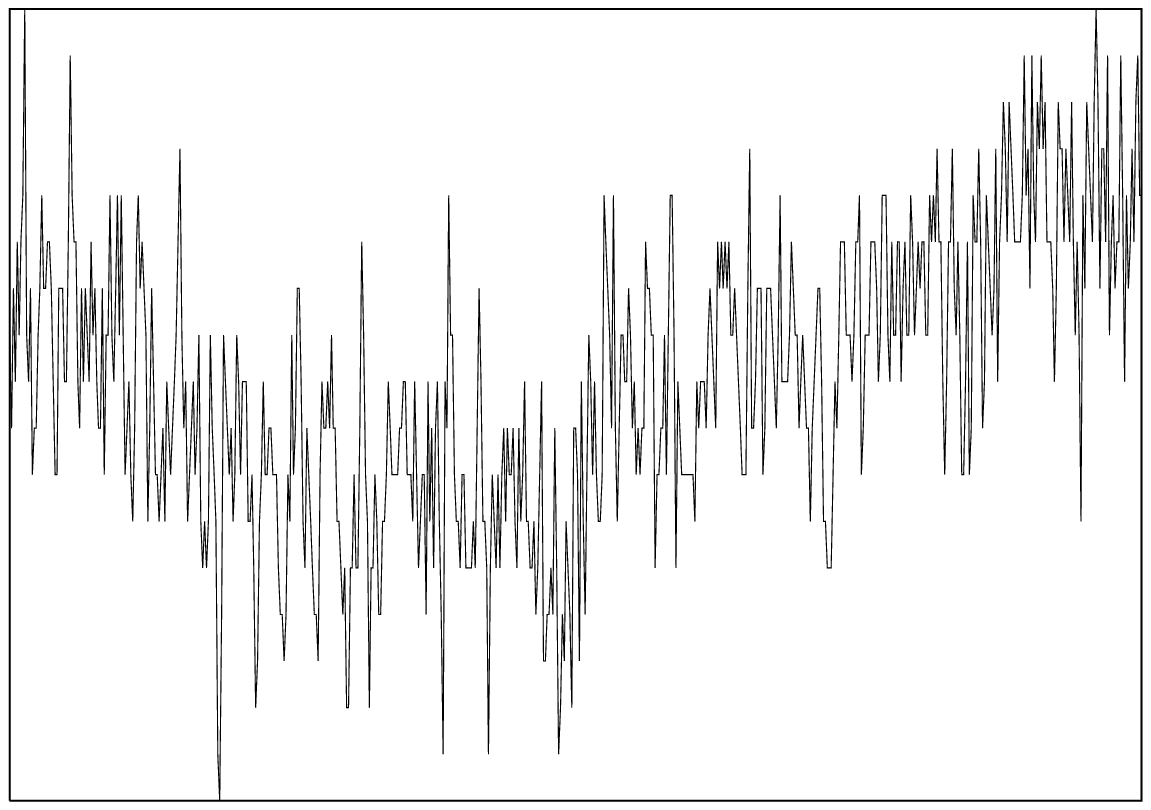}\\\includegraphics[width=7.75cm, 
height=1.5cm]{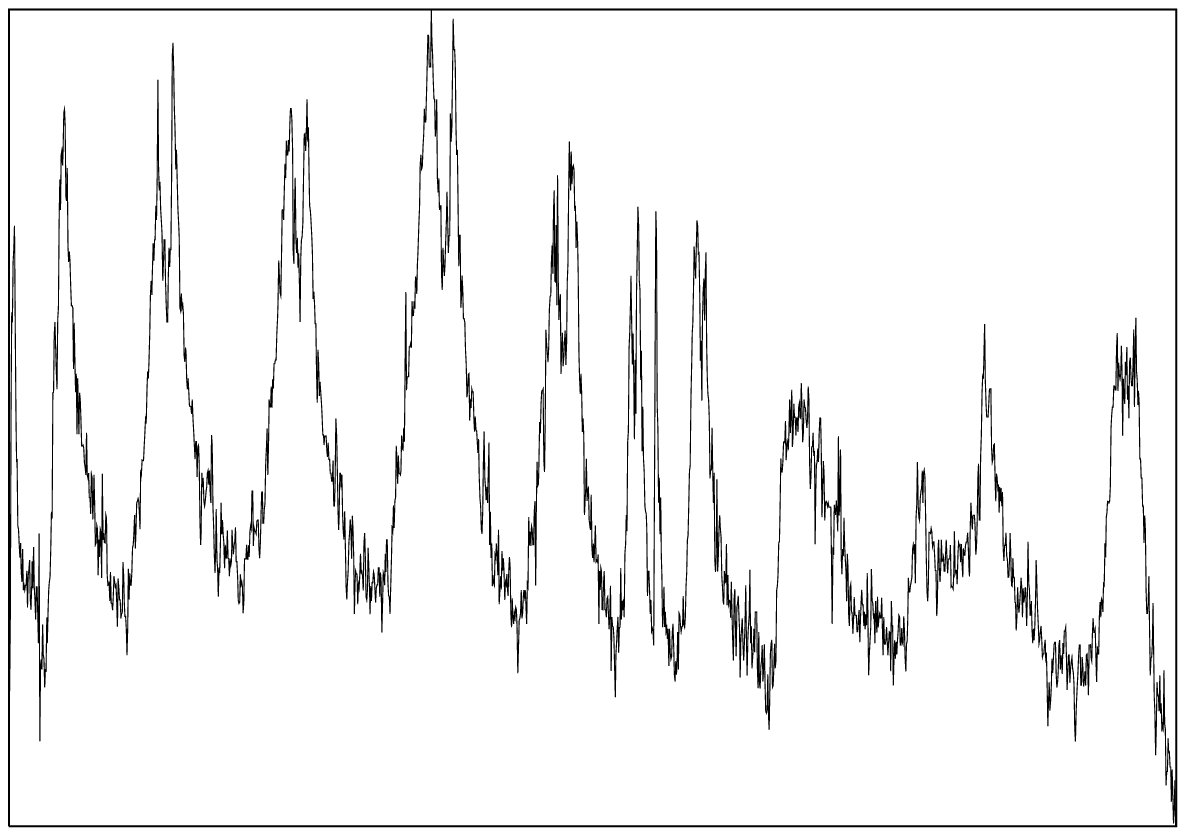}
\caption{Height of sand as a function of the position at a given Reynolds number: (Top) $R_e=10.4$, only fluctuations due to sand grains are visible. (Bottom) $R_e=34.6$, the instability is characterized by a periodic profile. The experimental parameters are: $\omega=45.18$ rpm and $h=1.6$ cm. Note that images have been rescaled in order to be compared.}\label{instab}
\end{center}
\end{figure}

We have performed experiments with values of $\omega$ ranging from 29 rpm to 71 rpm, and values of $h$ ranging from 1.05 cm to 3.36 cm. For each experiment, an image of the landscape has been taken. On each image, we have extracted transversal profiles of the landscape at 50 different $r$ values, i.e. different $R_e$ values. Figure \ref{instab} presents typical data for two values of the Reynolds number: $R_e=10.4$ ($r$=3 cm) and $R_e=34.6$ ($r$=10 cm). For small Reynolds number values, the profile is very noisy and no modulation is observed. The noise comes from the irregularities of the granular surface. On the contrary, we have noticed a periodic profile for large values of $R_e$. The transition between noisy and periodic regions is well defined and occurs at a given critical radius $r_c$. The profiles are periodic above $r_c$ and noisy elsewhere. This kind of behavior is characteristic of the emergence of an instability in the sand-water interface. This is similar to what is observed in the case of an instability in the boundary layer of a fluid flowing over a stainless-steel disk \cite{faller}. 

\begin{figure}[htb]
\begin{center}
\includegraphics[width=8.5cm]{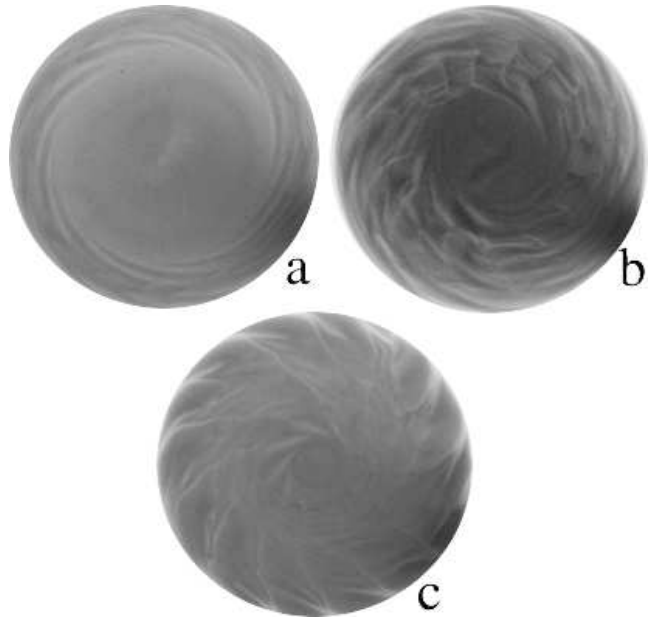}
\caption{Depending on the rotation speed, three patterns can be observed: one spiral, two superposing spirals and two paired spirals. (a) At low rotation speeds ($\omega\leq 35$ rpm) the ripples form one spiral. (b) At medium speeds ($35<\omega<45$ rpm) two spirals are superposed, one having arm azimuthal angles larger than the other ones. (c) At high rotation speeds ($\omega\geq 45$ rpm) the arms of the two spirals are coupled.}\label{phases}
\end{center}
\end{figure}

For all tested values of both parameters $h$ and $\omega$, we have noticed the appearance of that instability for high Reynolds number values. However, we have remarked that the pattern formed on the granular landscape depends on the experimental parameters. Indeed, we have recorded three kinds of patterns : (i) one spiral; (ii) two superposing spirals, one having the arms close to the tangential direction to the container circumference, the other being more open; and (ii) two paired spirals having the same number of arms and having their arms coupled one to the other. Figures \ref{phases}a-c illustrate those different patterns. 

\begin{figure*}[ht]
\begin{center}
\includegraphics[width=16cm, height=8.2cm]{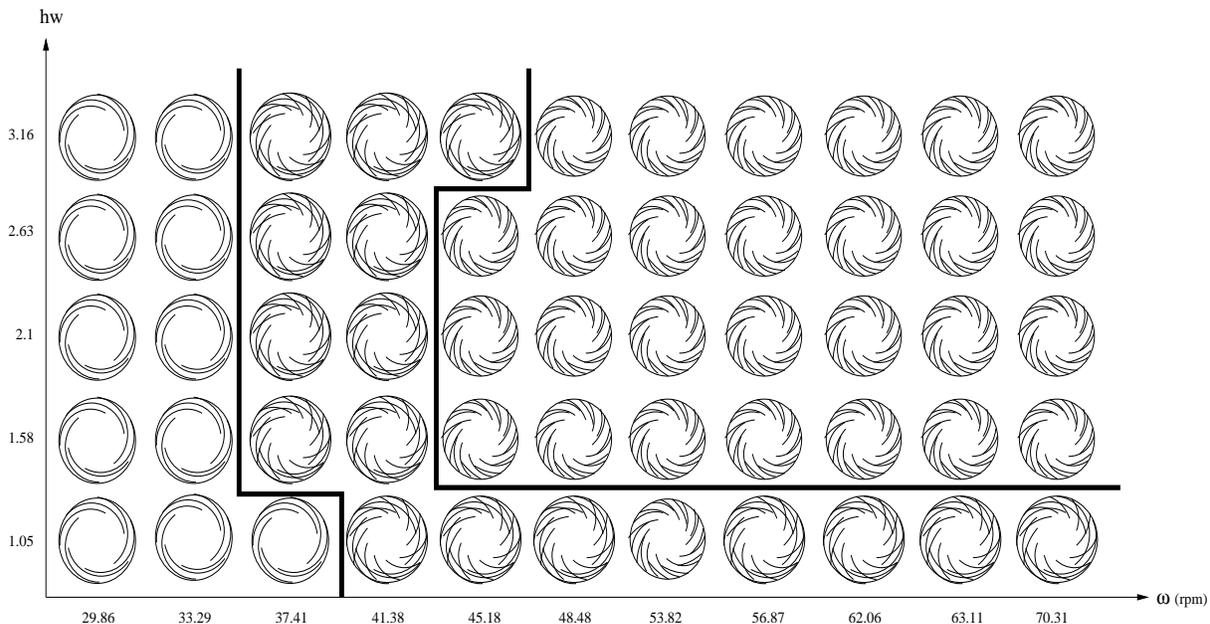}
\caption{Phase diagram giving the observed pattern as a function of the rotation speed $\omega$ and the height of water $h$. The rotation speed of the container clearly plays a role in the pattern formation, while the height of water has no significant effect. The three phases are separated by solid lines. At $\omega<35$ rpm, only one spiral is observed. A second spiral appears for $35<\omega<45$ rpm and is paired to the former spiral for $\omega>45$ rpm.}\label{diag_phase}
\end{center}
\end{figure*}

We have constructed a phase diagram giving the pattern as a function of both $\omega$ and $h$ parameters. One should note [see Figure \ref{diag_phase}] how the pattern depends on the rotation speed $\omega$ and is less dependent on the height of water $h$. Looking for details in images corresponding to the phase diagram, we have concluded that the patterns evolve as follows. At a low rotation speed ($\omega\leq 35$ rpm), spirals are created above a critical radius $r_c$. An increase of $\omega$ causes the increase of the spiral arm angles, as we will see further. At medium rotation speeds ($35<\omega<45$ rpm) a second spiral appears. The arms of this second spiral are nearly tangential to the cylinder circumference. As a consequence, one observes two superposed spirals. If the rotation speed is increased again ($\omega\geq 45$ rpm), the second spiral becomes more and more open, so that both spirals are paired. 

\subsection{Fourier analysis}

In order to determine the geometrical properties of the patterns, we have decomposed the data of the different profiles within Fourier series, according to the equation
\begin{equation}
f(x)=\frac{a_0}{2}+\sum_{m=1}^{N}\left[a_m\,\cos\left(\frac{2\pi 
mx}{L}\right)+b_m\,\sin\left(\frac{2\pi mx}{L}\right)\right],
\end{equation}
where $f(x)$ is the height of sand at position $x$ along the circle of radius $r$, $m$ is the Fourier mode, $a_m$ and $b_m$ are the Fourier coefficients. The sum is extended over $N$ terms, where $N$ is equals to the number of points in the discretization of $f(x)$. Typically, $N\approx 2000$. One should note that the geometry of our experimental setup allows us to use Fourier approximation without restriction. From this approximation, we can compute the statistical weight $A_m$ of a given mode $m$ in the series by
\begin{equation}
A_m=\sqrt{a_m^2+b_m^2}.
\end{equation}
The knowledge of all weights $A_m$ allows us to determine the mean amplitude \am of the ripples
\begin{equation}
\langle A\rangle=\sqrt{\sum_{m=1}^N A_m^2}.
\end{equation}
The mean wavelength of the patterns at the radius $r$ is then given by
\begin{equation}\label{mode_moyen}
\langle\lambda\rangle=2\pi r\left(\frac{\sum_{m=1}^N A_m\ m}{\sum_{m=1}^N A_m}\right)^{-1},
\end{equation}
which is the circumference at the considered radius $r$, divided by the statistical mean of the weighted modes $m$. It is important to note that practically, the sums are calculated from $m=4$ instead of $m=1$. In such a way, we remove the large modulations of the surface. Those large oscillations are artifacts due to the preparation of the sand bed before the experiment is started. Such problem is also encountered in other experiments dealing with granular surfaces \cite{frette,stegner}. 

\subsection{Critical radius $r_c$}

With the help of Fourier decompositions, one can easily find the value of $r_c$. For all radii $r<r_c$, the profiles are noisy. In such a noise, all frequencies have the same probability of being found in the profiles (white noise). Mathematically, this fact results in Fourier series which are sums of $sinus$ and $cosinus$ with nearly the same statistical weights $A_m$. On the contrary, for $r\geq r_c$ the periodic shape of the profiles implies the existence of a characteristic mode $M$ in each profile. The value of $M$ is equals to the number of ripples in the profile. The plot of $A_m$ as a function of the Fourier modes will thus exhibits a peak corresponding to the mode $M$. Eventually, the critical radius $r_c$ is equal to the radius where the first peak is recorded. Practically, $r_{c}$ is defined as the radius where the highest peak reaches a defined threshold value. 

\begin{figure}[htb]
\begin{center}
\includegraphics[width=8cm]{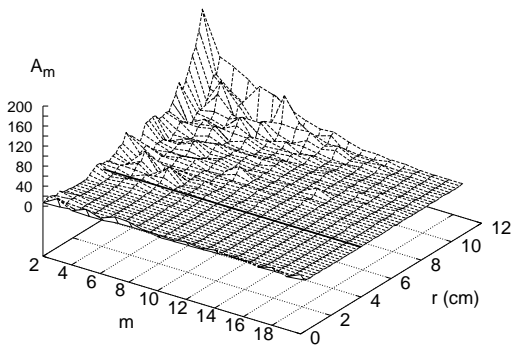}
\caption{Amplitude $A_m$ of the different Fourier modes $m$ as a function of the radius $r$. Since the instability is characterized by periodic profiles, Fourier peaks are observed in the spectrum only for radii $r>r_c$. On the contrary, stable regions have no periodic characteristics. The critical radius $r_c=4.1$ cm is emphasized by a solid line. The experimental parameters are: $\omega=45.18$ rpm and $h=1$ cm.}\label{modes}
\end{center}
\end{figure}

In Figure \ref{modes}, we report the amplitudes $A_m$ of the different Fourier modes as a function of the radius $r$. One can see that peaks appear only for $r\geq 4.1$ cm, meaning $r_c=4.1$ cm. Notice that in this case, $M\approx 6$ for $r=r_c$.

\begin{figure}[ht]
\begin{center}
\includegraphics[width=7.5cm]{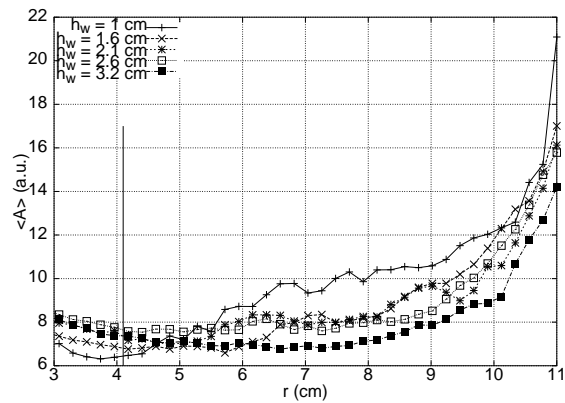}
\caption{Mean amplitude \am as a function of the radius $r$. 
Different values of the water height $h$ are illustrated. The solid vertical line gives the critical radius $r_c$. One should note that $r_c$ seems to be independent of $h$.}\label{rc}
\end{center}
\end{figure}

We have studied the dependence of $r_c$ on the height of water and the rotation speed. Figure \ref{rc} presents the mean ripple amplitude \am as a function of the radius $r$ for different values of $h$ at fixed $\omega=41.38$ rpm. First, one should note that the $r$-axis can be decomposed in 2 distinct regions: (i) If $r<r_c$, all curves are nearly constant. This is the stable part of the sand-water interface. (ii) For $r\geq r_c$, all curves grow. Notice that the critical radius is the same for all curves. This means that the birth of the instability is not controlled by the height of water. Moreover, one can see how \am is inversely proportional to $h$. Thus, the higher the velocity gradient of the water is, the larger are the ripples.

\begin{figure}[htb]
\begin{center}
\includegraphics[width=7.5cm]{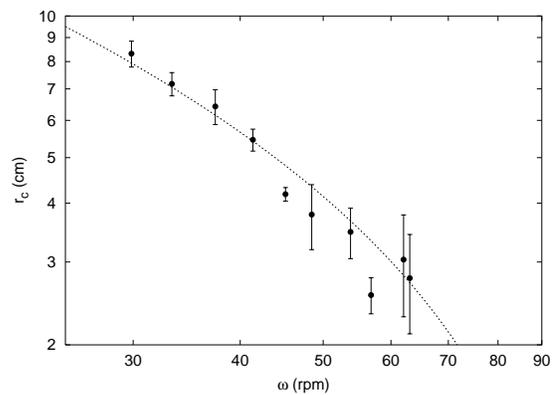}
\caption{Critical radius $r_c$ as a function of the rotation speed $\omega$. A fit using Eq.(\ref{power}) is also illustrated. The power exponent is found to be equal to -1/2.}\label{rc_w}
\end{center}
\end{figure}

Plotting the critical radius $r_c$ as a function of the rotation speed $\omega$ [see Figure \ref{rc_w}] shows that the instability rises as sooner as the fluid speed is high. Fitting the data with the single power law
\begin{equation}\label{power}
r_c(\omega)=a+b\, \omega^\tau,
\end{equation}
we found that $\tau$ is roughly equal to -1/2. From the definition Eq.(\ref{reynolds}), we deduce that the {\it critical} Reynolds number $R_{e_c}$ is constant and is
\begin{equation}
R_{e_c}=\sqrt{r_c^2\frac{\omega}{\nu}}\sim 16.
\end{equation}

\subsection{Ripple wavelength \lm}

Using Eq.(\ref{mode_moyen}), we have studied the evolution of the mean ripple wavelength \lm from the center of the plate, to the sides of the container. Below the critical radius $r_c$, no modulation is observed and \lm is nearly zero. 

Within the unstable part of the interface, the measurement of the mean ripple wavelength is perturbated by the presence of ``defects''. Indeed, when two ripples collapse a jump of \lm is recorded. We have noticed that on an image, the defects are almost situated at the same radius. Their role is to adjust the number of ripples to the shear conditions. In order to increase the accuracy of the measurements, we have decomposed the whole $r$-range into intervals which do not contain such ``defects''.

\begin{figure}[htb]
\begin{center}
\includegraphics[width=7.5cm]{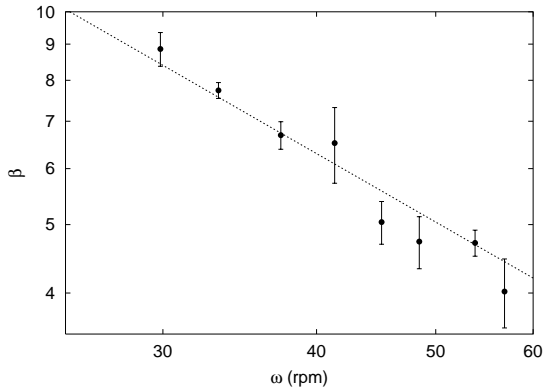}
\caption{Log-log graph giving the growth rate $\beta$ of the mean ripple wavelength \lm as a function of the rotation speed $\omega$. A fit using Eq.(\ref{lambda_w}) is also illustrated.}\label{b_w}
\end{center}
\end{figure}

We have found that \lm grows linearly with the distance form the center of the plate:
\begin{equation}
\langle\lambda\rangle=\beta\ r\quad \text{for}\ r>r_c.
\end{equation}
This observation is consistent with the existence of spirals with constant arm numbers in the considered intervals. Moreover, we have noticed that the growth rate $\beta$ does not depend on the height of water but decreases with the rotation speed $\omega$, according to a power law with power exponent -1. Eventually, we see that the mean ripple wavelength obeys the scaling
\begin{equation}\label{lambda_w}
\langle\lambda\rangle\sim\frac{r}{\omega} \quad \text{for}\ r>r_c.
\end{equation}
In Figure \ref{b_w}, we have plotted $\beta$ as a function of $\omega$ as well as a fit using Eq.(\ref{lambda_w}).

An interpretation of the decrease of $\beta$ can be deduced from the phase diagram illustrated in Fig.\ref{diag_phase}. As the height of water is changed at fixed rotation speed, no major modification of the patterns is observed (nearly same wavelength). The consequence is that $\beta$ does not depend on $h$. On the other way, we have seen that $\omega$ may cause the appearance of a second spiral. Such an occurrence results in a larger amount of spiral arms, i.e. in a larger number of ripples. Therefore, the mean ripple wavelength is lower when $\omega$ increases. The small values of $\beta$ at large $\omega$ values clearly shows this fact.

An important observation is that {\it no well-defined transition} is observed in the curve giving \lm as a function of $\omega$. Indeed, the decrease of the mean ripple wavelength seems to be continuous. The way of appearance of the so-called second spiral is one reason explaining this result. At $\omega\approx 35$ rpm, the second spiral is formed but has a smaller amplitude than the former spiral. The statistical weight of its wavelength is thus smaller than those of the former spiral. As the rotation speed increases, the second spiral amplitude growth. Eventually, the measured mean ripple wavelength corresponds to the number of spiral arms of both spirals, i.e. becomes at least twice those measured when only one spiral was observed. One should note that since \lm is divided by a factor larger than $2$, we see 
that the hydrodynamics also tends to increase the number of arms of each spiral. This has been mentioned here-above as the origin of the ``defects'' in the spiral patterns.

\subsection{Azimuthal angle \az}

\begin{figure}[h]
\begin{center}
\includegraphics[width=6cm]{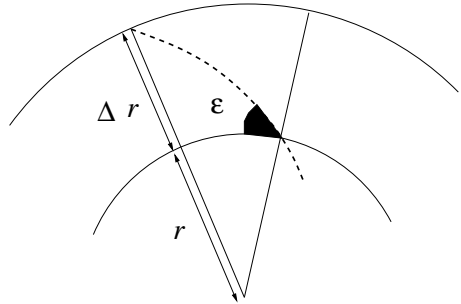}
\caption{The azimuthal angle \az is defined as the angle between the ripples crest (dashed line) and the tangential direction. The azimuthal angle \az is measured between two concentric circles (solid lines) of radii $r$ and $r+\Delta r$ respectively ($\Delta r=0.25$ cm)}\label{mesure_epsilon}
\end{center}
\end{figure}

Going deeper into the geometrical description of the patterns, we have measured the azimuthal angle \az of the spiral arms. This parameter is defined at a given point by the angle between a ripple crest and the tangential line to the circle centered on the plate and passing by the considered point [see Fig.\ref{mesure_epsilon}]. Practically, \az is measured between two concentric circles with radii differing from $\Delta r=0.25$ cm.

In order to determine the influence of the experimental parameters $h$, $\omega$ and $r$ on the azimuthal angle \az, we have proceeded as follows. At a given distance from the center of the container, i.e. a given $r$-value, and a given $\omega$-value, we measured \az for different values of the height of water $h$. This process is repeated all over the container diameter. Then, the height of water $h$ was fixed, and the procedure explained above was applied for different values of the rotation speed $\omega$. 

\begin{figure}[hbt]
\begin{center}
\includegraphics[width=7.5cm]{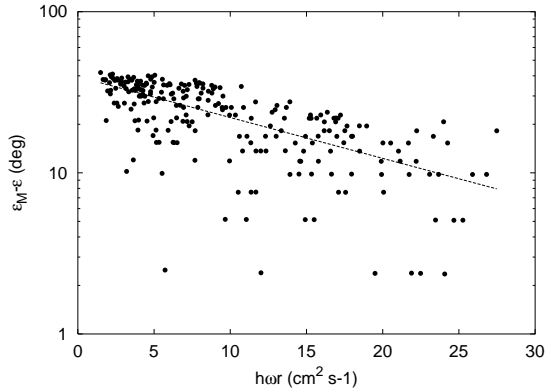}
\caption{Semi-log graph of the azimuthal angle \az as a function of the rotation speed $\omega$ multiplied by the height of water $h$ and the radial distance $r$. A fit using Eq.(\ref{eqaz}) is also illustrated.}\label{e_w}
\end{center}
\end{figure}

We have noticed that the azimuthal angle \az increases with the distance from the center of the container as well as with the height of water $h$ and the rotation speed $\omega$. Moreover, the measurements collapse when they are plotted as a function of the product $h\omega r$ [see Fig.\ref{e_w}]. It appears that the experimental data could be considered as following a power law as well as saturating exponentially. We have chosen the exponential law in order to avoid the azimuthal angle to grow to infinity. The fit of the data sets shows that 
\begin{equation}\label{eqaz}
\varepsilon=\varepsilon_M-\alpha_1\exp{\left(-\frac{h\omega r}{\alpha_2}\right)},
\end{equation}
where $\varepsilon_{M}=47^{\circ}\pm 8^{\circ}$ is the asymptotical value of the azimuthal angle at high $h\omega r$ values, $\alpha_1$ and $\alpha_2$ being fitting parameters.

From Eq.(\ref{eqaz}), one can see that the azimuthal angle \az meets the horizontal axis, says \az=0, for the critical $h\omega r$ product value $p_{c}$ 
\begin{equation}\label{vc}
p_c=\alpha_2\,\ln\left(\frac{\varepsilon_M}{\alpha_1}\right)=3\, 10^{-4}\text{\ m}^{2}\text{\ s}^{-1}.
\end{equation}
Since the azimuthal angle is zero at this critical value $p_c$, concentric circles must be observed for $h\omega r=p_c$ and spirals with azimuthal angles $\varepsilon<0$ for values of $h\omega r<p_c$. However, we have never observed such patterns. 

An interpretation of this can be deduced from the consideration of the fluid momentum $P=mv$, where $m$ is the mass of fluid, and $v$ its linear speed. Considering that $\rho=1\, 10^3$ kg m$^{-3}$ is the density of the fluid, and that the radius of the container $R=0.11$ m, one gets
\begin{equation}\label{eqp}
P=\rho\pi R^2\, h\omega r,
\end{equation}
what contains the $h\omega r$ term. This thus means that the value of the azimuthal angle \az is governed by the fluid momentum, according to Eq.(\ref{eqaz}). Combining Eq.(\ref{eqaz}) and Eq.(\ref{eqp}) we are able to predict the fluid momentum corresponding to concentric circles. This critical momentum $P_c$ reads
\begin{equation}\label{eqp2}
P_c=\rho\pi R^2\, p_c=0.011\text{\ kg}\text{\ m}\text{\ s}^{-1}.
\end{equation}
For fluid momentum $P<P_c$, spirals with negative azimuthal angles must be created. 

The fact that no concentric circles are observed comes from the lack of granular transport for such small fluid momentum. Indeed, ripples are only created when the grains are transported in the fluid, and this is only possible if the fluid shear stress exceeds a certain threshold. Typically, ripples are observed only for fluid momentum $P>0.1$ kg m s$^{-1}$.

\section{Discussion}

The instability encountered in our experiment creates patterns very similar to those observed in the case of a fluid flowing over a disk at rest \cite{jarre,faller,hoffmann,gauthier}. The instability resulting in such patterns takes place in the B\"odewadt boundary layer, i.e. in the boundary layer over the bottom disk. The patterns we observe are also similar to those observed in the Thomas' experiment \cite{thomas}. In the latter experiment the ripples are created by an acceleration of the container. The acceleration rate $\Delta\omega$ is somewhat analogue to the rotation speed $\omega$ presented here. Hereafter, our experiment and the Thomas one will be referred as ``granular'' one, while \cite{jarre,faller,hoffmann,gauthier} will be defined as ``fluid'' experiments.

Despite the similarity between the patterns, it is not obvious that the spiral patterns observed here are of the same nature as the fluid ones or as the Thomas ones. We are now attempting to make some comparison in order to differentiate the different kinds of instabilities. It is important to notice that B\"odewadt flows are very complicated to create experimentally. As a consequence, there is a poor amount of data.

The major difference between the ``granular experiments'' and the fluid ones is the dependence of the mean wavelength of the patterns on the distance from the center [see Table \ref{order}]. In Thomas and our experiments, the mean pattern wavelength increases with distance from the center while it is nearly constant in fluid experiments. However, the order of magnitude of \lm for all experiments is about 1 cm.

The values of the azimuthal angle of the spirals observed in fluid experiments range between $13^\circ$ and $17^\circ$, what is nearly constant compared to the angles measured by Thomas (between $20^\circ$ to $70^\circ$) and the angles we measured (from $4^\circ$ at low fluid momentum to $45^\circ$ at higher fluid momentum). The most important observation is that \az increases with the radius in our experiment, while it decreases in \cite{thomas}. 

Measurements of critical Reynolds numbers show that $R_{e_c}$ decreases with the rotation speed in all experiments. However, the dependence is very slight in fluid experiments, compared to what is observed in granular ones. Moreover, $R_c\approx40$ in fluid experiments and is only around $16$ here.

\begin{center}
Scaling\\
\begin{tabular}{c | c c c}
\hline\hline
Parameters & Here & Thomas\cite{thomas} & B\"odewadt\cite{schlichting}\\
\hline
\lm  & $\sim r/\omega$ & $\sim r$ & $\sim cte$ \\
\az &  $\sim h\omega r$ & $\sim r^{-0.5}$ & $\sim cte$\\
$R_{e_c}$ & $\approx 16$ & $\sim \omega^{-0.1}$ & $\approx 40$ \\
\hline\hline
\end{tabular}
\end{center}
\begin{table}[h]
\caption{Behaviors of the physical parameters \lm, \az and $R_{e_c}$ for 
different experiments.}\label{order}
\end{table}

From the comparison of the orders of magnitude and the behaviors, we can conclude that the instability observed here is not the signature of an instability in the boundary layer. What we observed here are ripples, similar to those observed in coastal areas, but in a rotating frame. Thomas pointed out the same conclusions. In both cases the spiral shape formed by the ripple crests may be due to the presence of centrifugal forces and other effects caused by the motion of the grains in the fluid. However, due to the differences between the experimental procedures followed in \cite{thomas} and here, the created patterns do not behave in the same way. 

\section{Summary}
An experimental study of the instability rising on a sand-water interface has been proposed. The formation of the spiral patterns have been shown and have been described with the help of a phase diagram.

The properties of the critical radius where the instability forms have been studied. The critical radius $r_c$ is found to decrease with the rotation speed of the container, according a square law, giving a critical Reynolds number equals to $R_{e_c}\approx 16$.

The mean ripples wavelength is proportional to the distance from the center of the container and inversely proportional to the rotation speed of the container. Studying the azimuthal angles of the spirals, we have shown that this angle is governed by the momentum of the fluid $P=mv$, where $m$ is the mass of fluid, and $v$ its linear speed. From fits of experimental data, the azimuthal angle is found to saturates at around $45^\circ$ for very high fluid momentum. A comparison between instabilities rising in the B\"odewadt boundary layer, as well as a comparison with an experiment of ripple formation in an accelerated frame, allows us to conclude that the observed patterns are not the signature of a fluid instability. The instability of the sand-water interface we observe are similar to natural ripples created in a rotating frame. 

\section*{Acknowledgements}
HC is financially supported by the FRIA, Belgium. This work is also supported by the ARC contract of the University of Li\`ege. The authors would like to thank F. Zoueshtiagh for fruitful discussions.


\begin{thebibliography}{3}
\bibitem{jarre} S. Jarre, P. Le Gal and M. P. Chauve, {\em Europhys. Lett.} {\bf 14}, 649 (1991).
\bibitem{faller} A. J. Faller, {\em J. Fluid. Mech.} {\bf 230}, 245 (1991).
\bibitem{hoffmann} N. Hoffmann, F. H. Busse and W.-L. Chen, {\em J. Fluid. Mech.} {\bf 366}, 311 (1998).
\bibitem{gauthier} G. Gauthier, P. Gondret and M. Rabaud, {\em J. Fluid. Mech.} {\bf 386}, 105 (1999).
\bibitem{schlichting} H. Schlichting, {\em Boundary Layer Theory}, (Mc Graw-Hill, NY, 1962). 
\bibitem{frette} A. Betat, V. Frete, and I. Rheberg, {\em Phys. Rev. Lett.} {\bf 83}, 88 (1999).
\bibitem{stegner} A. Stegner and J. E. Wesfreid, {\em Phys. Rev. E} {\bf 60}, R3487 (1999).
\bibitem{makse} H. A. Makse, {\em Eur. Phys. J. E} {\bf 1}, 127 (2000).
\bibitem{thomas} P. J. Thomas, {\em J. Fluid. Mech.} {\bf 274}, 23 (1994).
\bibitem{farzam} F. Zoueshtiagh and P. J. Thomas, {\em Phys. Rev. E} {\bf 61}, 5588 (2000).
\bibitem{bagnold1} R. A. Bagnold, {\em The physics of blown sand and desert dunes}, (Chapman and Hall, London, 1941). 
\end{thebibliography}
\end{document}